\documentclass[twocolumn,aps,prl,showpacs]{revtex4}
\usepackage{amsmath} 
\usepackage{amssymb} 
\usepackage{graphicx} 
\usepackage{epsfig}
\usepackage{epstopdf}
\usepackage{color}
\usepackage{rotating}
\usepackage{bm}
\usepackage{xspace}
\newcommand{\Mo}{Mo$_3$S$_7$(dmit)$_3$\xspace}
 
\begin{document}
 \title{Topological quantum phase transition driven by anisotropic spin-orbit coupling in trinuclear organometallic coordination crystals}
 \author{J. Merino}
\affiliation{Departamento de F\'isica Te\'orica de la Materia Condensada, Condensed Matter Physics Center (IFIMAC) and
Instituto Nicol\'as Cabrera, Universidad Aut\'onoma de Madrid, Madrid 28049, Spain}
  \author{A. C. Jacko}
 \affiliation{School of Mathematics and Physics, The University of Queensland,
Brisbane, Queensland 4072, Australia}
\author{A. L.  Khosla}
\affiliation{School of Mathematics and Physics, The University of Queensland,
	Brisbane, Queensland 4072, Australia}
 \author{B. J. Powell}
 \affiliation{School of Mathematics and Physics, The University of Queensland,
Brisbane, Queensland 4072, Australia}
 \begin{abstract}
We show how quasi-one-dimensional correlated insulating states arise at two-thirds filling in organometallic multinuclear coordination complexes described by layered decorated honeycomb lattices.
The interplay of spin-orbit coupling and electronic correlations leads to pseudospin-1 moments arranged in
weakly coupled chains with highly anisotropic exchange and a large trigonal splitting. This leads to a  quantum phase transition from a Haldane phase to a topologically trivial phase as the relative strength of the spin-orbit coupling increases.
 \end{abstract}
 \date{\today}
\maketitle
 
 {\em Introduction.}
In recent years there has been intense research activity on the effect of spin orbit coupling (SOC)
in the electronic structure of weakly interacting materials since the discovery of topological band insulators \cite{hasan2010,zhang2011} (TBI).
TBI are predicted to occur in certain solids with electron-like quasiparticles in the presence of strong SOC  as is found in 
compounds containing the heavy elements such as: Bi, Pb, Sb, Hg and Te.
In materials with partially filled localized orbitals, renormalization effects induced by strong Coulomb  
repulsion effectively enhance the SOC leading, for instance, to topological Mott \cite{pesin2010,balents2014} or Kondo \cite{Galitski} insulators. In Ir-based compounds 
such as Na$_2$IrO$_3$ and Li$_2$IrO$_3$ the emergent $J=1/2$ pseudospins arranged 
in a honeycomb lattice interact through anisotropic spin exchange couplings  of the Kitaev-Heisenberg type \cite{jackeli2009,perkins2014a,perkins2014b,Kee}.
Kitaev's honeycomb model can be solved exactly  \cite{kitaev2006} sustaining a topological quantum spin liquid state with Majorana fermion excitations
which may have been recently observed in the Kitaev candidate material  \cite{banerjee2016} $\alpha$-RuCl$_3$.
This illustrates how novel quantum phases of matter arise from the interplay of strong Coulomb repulsion and SOC in certain materials.

\begin{figure}%[h]
   \centering 
 \includegraphics[width=4.cm]{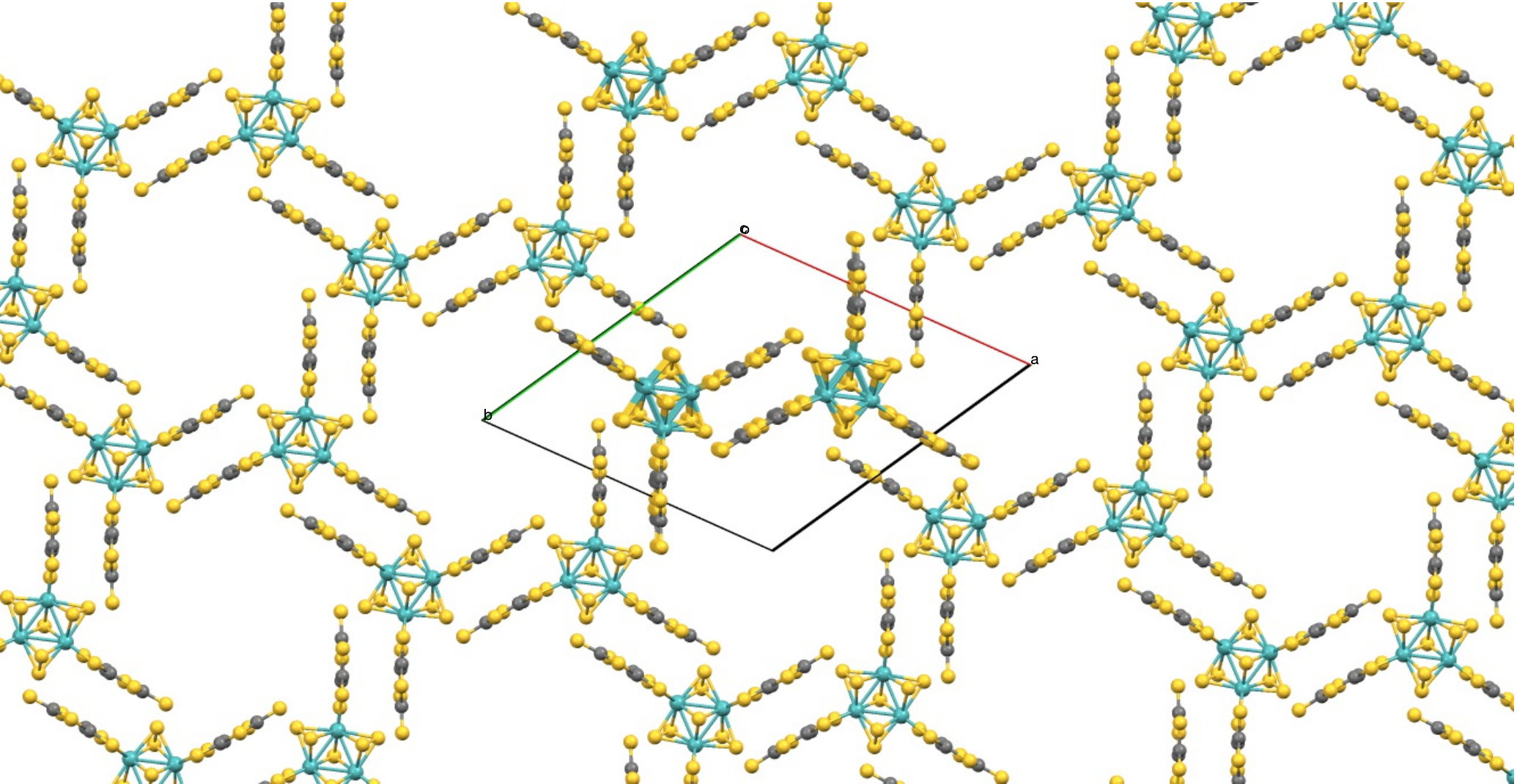}
  \includegraphics[width=4.4cm]{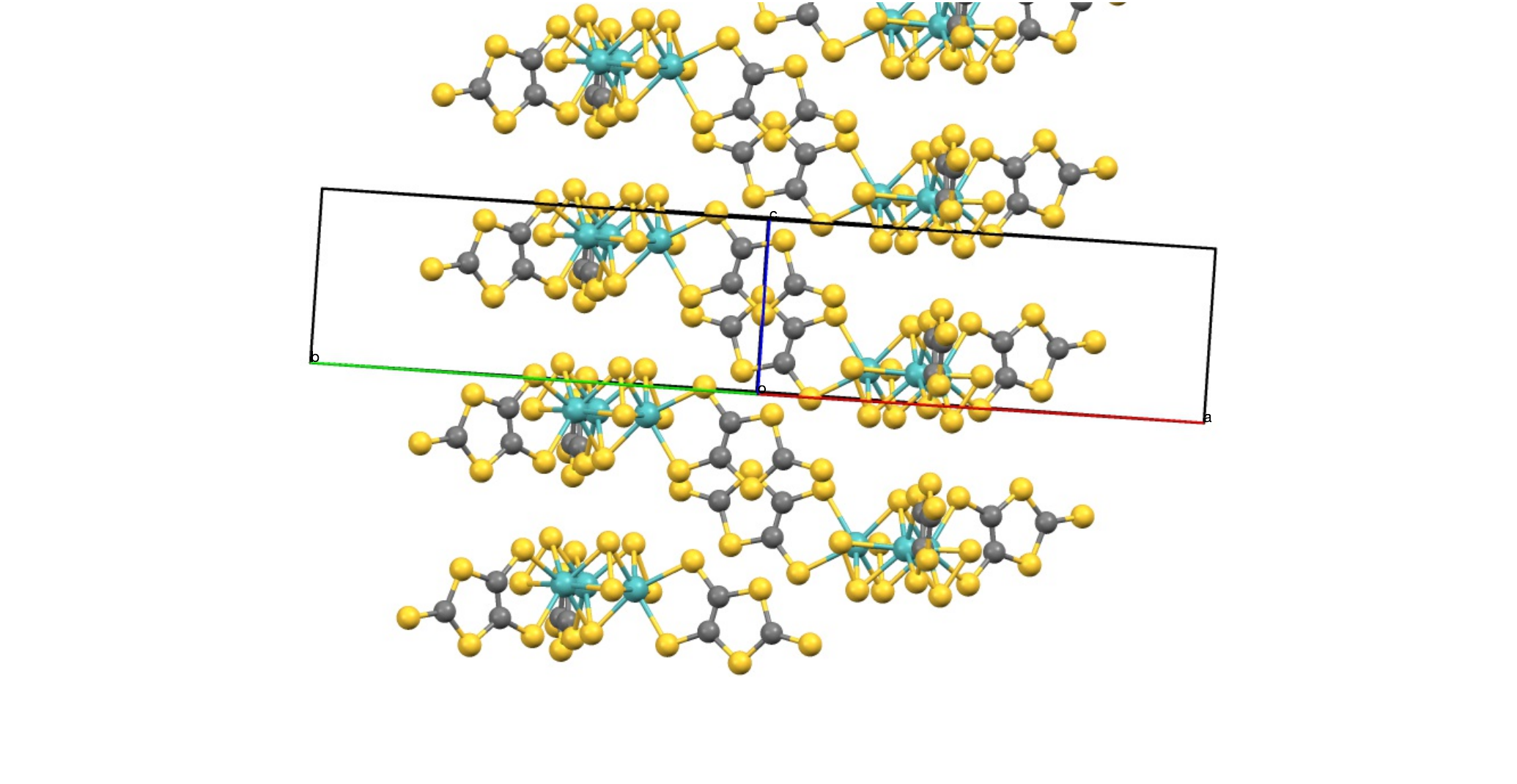}
   \caption{(Color online) Crystal structure of Mo$_3$S$_7$(dmit)$_3$. a) In the $a$-$b$ plane the triangular molecules form a honeycomb lattice. As each complex contains three Wannier orbitals this results in a decorated lattice, which extrapolates between the Kagome and graphene lattices \cite{jacko2015}. b) A complexes stack in chains along the $c$-axis resulting in a triangular tube arrangement of the Wannier orbitals.}
 \label{fig:fig1}
 \end{figure}

Molecular materials are ideal playgrounds to explore strong electronic correlation effects in low dimensions \cite{kanoda2011,powell2011}.  
Although SOC effects are generally weak in organic systems \cite{winter2012}, organometallic complexes provide a route for enhancing SOC \cite{powell2015}. Llusar {\it et al.} have synthesized a particularly interesting family of trinuclear organometallic  coordination  complexes \cite{guschin2013,llusar2010} with ligands that facilitate electronic transport between molecules including \Mo, Mo$_3$Se$_7$(dmit)$_3$, Mo$_3$S$_7$(dsit)$_3$, and Mo$_3$Se$_7$(dsit)$_3$ (dmit=S$_5$C$_3$, dsit=Se$_2$S$_3$C$_3$). For example, the low-energy electronic structure of  Mo$_3$S$_7$(dmit)$_3$ crystals \cite{llusar2004} is described by three Wannier orbitals per molecule that are hybrids of the Mo $d$-orbitals and the dmit molecular orbitals leading 
to narrow band structures with appreciable SOC \cite{jacko2015,Amie}. 
The Mo atoms, and hence the Wannier orbitals, form a layered structure with a decorated honeycomb lattice in the basal plane \cite{jacko2015}, as shown in Fig. \ref{fig:fig1}. In the $c$-direction, the triangular 
Mo$_3$S$_7$(dmit)$_3$  complexes arrange in tubes reminiscent of the CrAs tubes formed in the recently
discovered superconductor K$_2$Cr$_3$As$_3$ \cite{bao2015,zhong2015}. %The electronic structure of  a single Mo$_3$S$_7$(dmit)$_3$  molecule consists of an A orbital and a doubly degenerate E$^\pm$ orbitals reflecting the underlying trigonal ($C_3$) symmetry of the complex. 

At two-thirds filling in the presence of strong Coulomb
repulsion and no SOC, a correlated insulator consisting of spin-1 effective moments coupled through an 
isotropic antiferromagnetic interaction arises \cite{janani2014a,janani2014b,nourse}.
In this Letter, we derive the
effective superexchange model for trinuclear crystals including SOC.   SOC entangles the spin and orbital degrees 
of freedom  leading to pseudospin-1 moments, ${\bm{\mathcal{ S}}}_m$, which interact via anisotropic exchange interactions. 
The SOC also introduces a trigonal splitting of the triplet into a lower energy ${{\mathcal{ S}}}_m^z=0$ state
and a ${{\mathcal{ S}}}_m^z=\pm1$ doubly degenerate state. 
Surprisingly, we also find that the effective dimensionality of the pseudospin model is strongly affected by the strength of the electronic correlations: with increasing Hubbard $U$ the pseudospin-1 model becomes increasingly one-dimensional, even though the original crystal is almost isotropic. 
At large trigonal splitting, a `$D$-phase' consisting of the tensor product of ${{\mathcal{ S}}}_m^z=0$ states on each complex, occurs.
Thus, our analysis suggests that a topological quantum phase transition from a Haldane phase to a
topologically trivial $D$-phase can be induced by increasing the SOC.

%It is important to understand whether such anisotropic
%spin models on the lattices  of Fig. \ref{fig:fig0} can sustain novel spin liquid states.    

 {\em Pseudospin-1 moments in trinuclear complexes.}  We first analyze the local effective spin degrees of freedom
 arising at each isolated trimer due to the combined effect of Coulomb repulsion and SOC. %The low-energy physics of a single Mo$_3$S$_7$(dmit)$_3$ complex can be understood in terms of three Wannier orbitals, which are hybrids of the  Mo $d$-orbitals and (dmit)$_3$ ligand molecular orbitals  \cite{jacko2015,Amie}.
 We construct our model based on the Wannier orbitals found in recent density functional calculations \cite{jacko2015,Amie} and we model each complex via  a Hubbard-Heisenberg model 
 on a triangle with one orbital per site. 
 In molecules with cyclic symmetry, such as the $C_3$ symmetry  of a Mo$_3$S$_7$(dmit)$_3$ complex, the leading SOC term in the effective low-energy Hamiltonian couples the electron 
 spin to currents running around the complex  \cite{Amie}, which carry a `molecular' angular momentum, $\bm L_m$. For molecules with $N$-fold rotation axes the angular momentum carried by this current is $l=(N-1)/2$ if $N$ is odd \cite{Amie};  for a  Mo$_3$S$_7$(dmit)$_3$ complex $N=3$ and $l=1$.
 Thus  the Hamiltonian for the $m$th complex is $H_{c}=H_0+H_{int}+H_{SO}$, 
where
  \begin{eqnarray}
H_0&=&-t_c \sum_{\langle ij\rangle \sigma}\left(c^\dagger_{mi\sigma} c_{mj\sigma} + H. c. \right)
   \\\nonumber 
 H_{int}&=&U\sum_i n_{mi\uparrow}n_{mi\downarrow} +J_F \sum_{\langle ij \rangle} \left({\bf S}_{mi}\cdot {\bf S}_{mj} -{n_{mi} n_{mj} \over 4}\right) 
  \\
  H_{SO}&=& \lambda_{xy}(L_m^xS_m^x+L_m^yS_m^y)+\lambda_z L_m^zS_m^z,\nonumber
  \end{eqnarray}
   $c^{(\dagger)}_{mi\sigma}$ annihilates (creates) an electron  with spin  $\sigma$ in the  $i$th Wannier orbital of the $m$th molecule, 
  ${\bf S}_{mi}=\sum_{\sigma\sigma'}c^{\dagger}_{mi\sigma}{\bm\tau}_{\sigma\sigma'}c_{mi\sigma'}$, 
  $\bm\tau$ is the vector of Pauli matrices, 
  ${\bf S}_{m}=\sum_i{\bf S}_{mi}$, 
  $n_{mi\sigma}=c^{\dagger}_{mi\sigma}c_{mi\sigma}$,
  $n_{mi}=\sum_\sigma n_{mi\sigma}$,
   $t_c$ is the intramolecular hopping amplitude, 
  $U$ is the onsite Coulomb repulsion, and
  $J_F<0$ is the direct intramolecular ferromagnetic exchange  \cite{Fazekas}.  We take the  $z$-direction
  along the $c$-direction of the crystal.
  We consider the  three-site cluster with four electrons, relevant to Mo$_3$S$_7$(dmit)$_3$ and its selenated analogs.

\begin{figure}%[h]
   \centering
    \includegraphics[width=4cm]{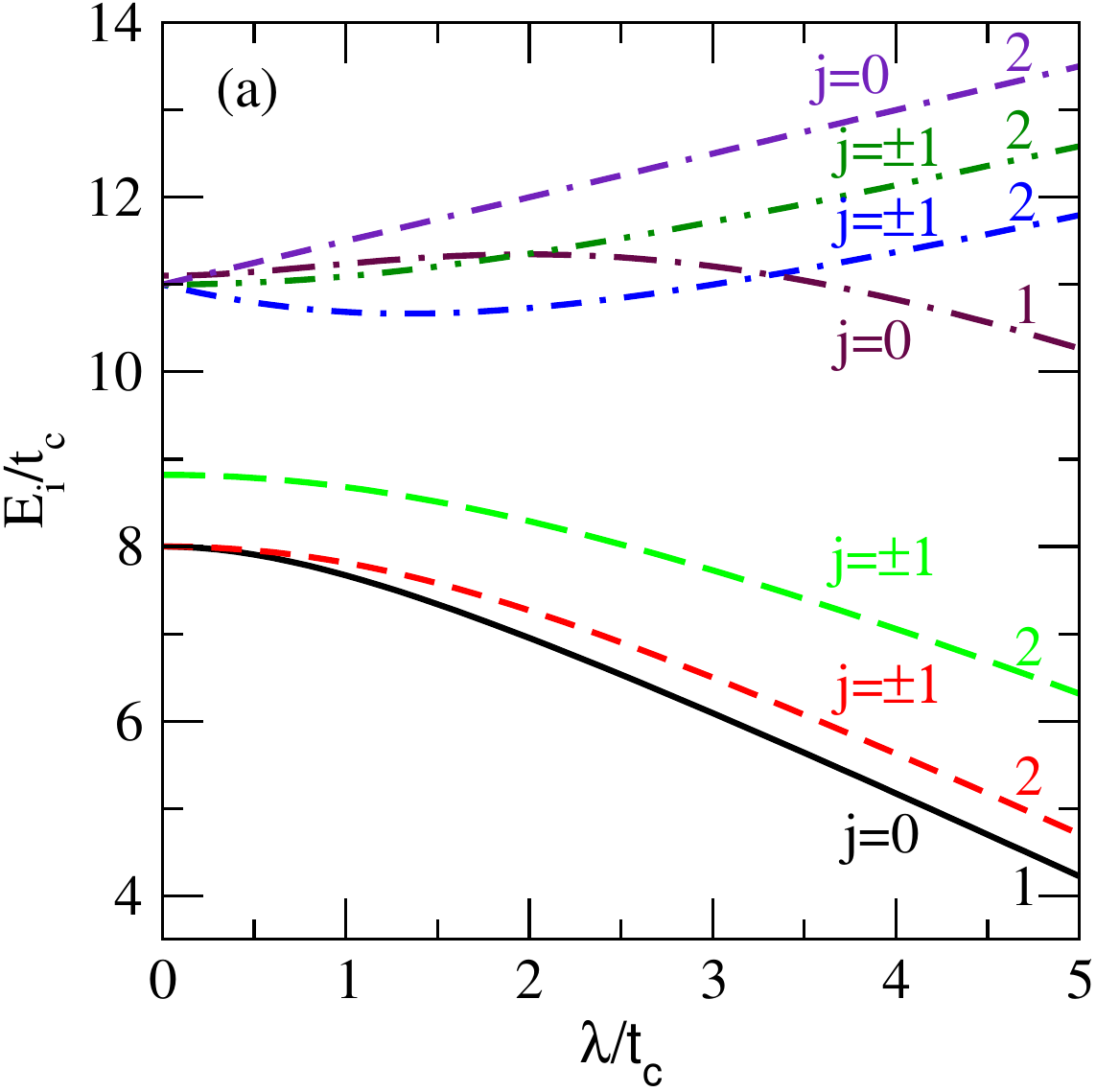}
     \includegraphics[width=4cm]{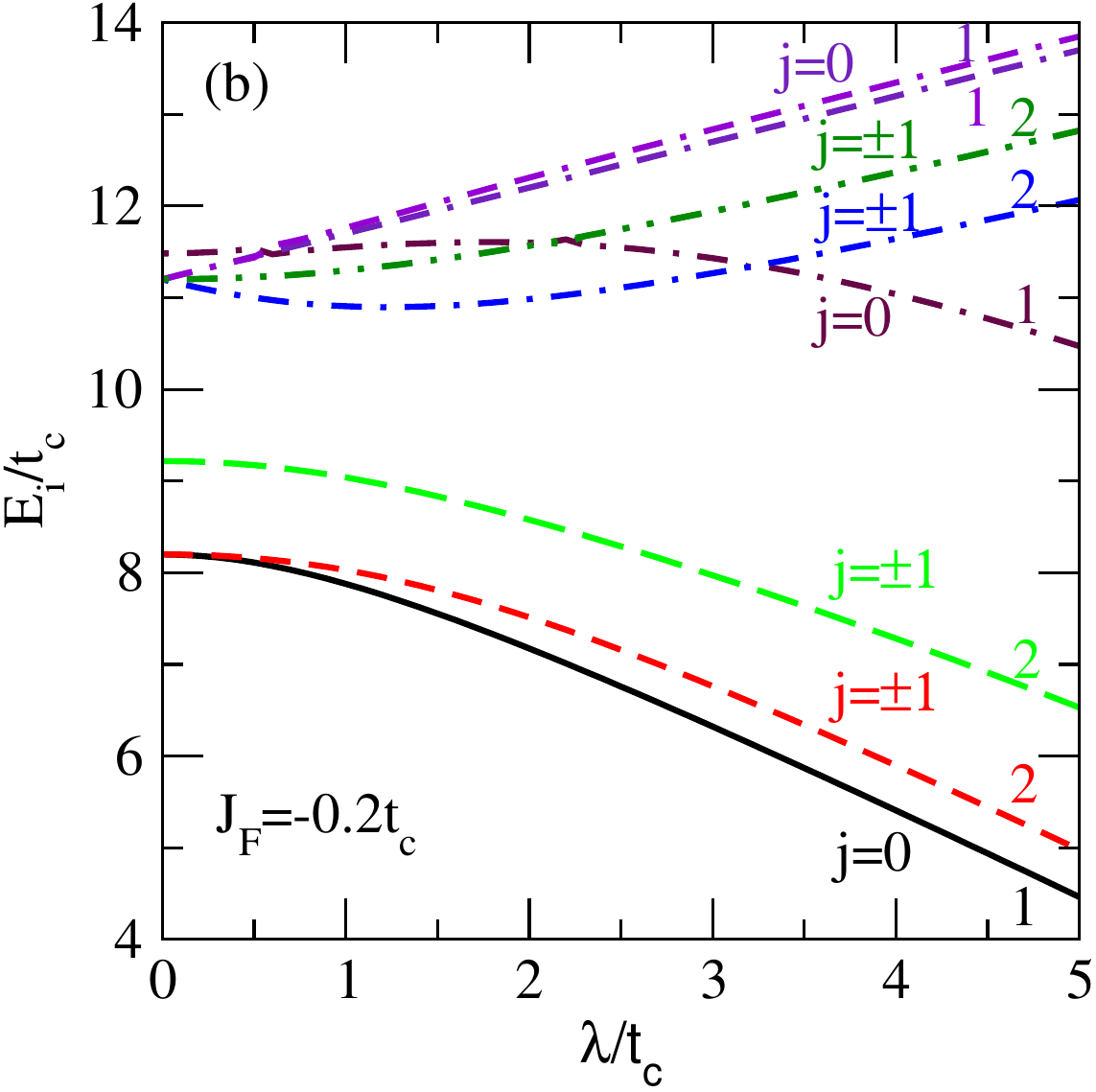}
   \caption{(Color online) Effective pseudospin-1 moments arising in isolated trimers in the presence of SOC. The   
   lowest-lying  eigenstates of the Hubbard-Heisenberg model describing trinuclear complexes with four electrons are shown
 for the pure Hubbard model ($U=10t_c, J_F=0$) (a) and for non-zero intracluster exchange (b) ($U=10t_c,J_F=-0.2t_c$).
 Only the $z$-component of the total angular momentum, $j$, is conserved labels. The
 degeneracies of the eigenstates are labelled by the numbers on the right-hand side of the figure.
As SOC, $\lambda=\lambda_{xy}=\lambda_z$, increases the triplet state is 
split into a $j=0$ state and a doubly degenerate $j=\pm1$ state. This is described 
by a pseudospin-1 in the presence of a trigonal field generated by SOC.
  }
 \label{fig:fig2}
 \end{figure}

\begin{figure}%[h]
 \includegraphics[width=4.25cm, clip=true]{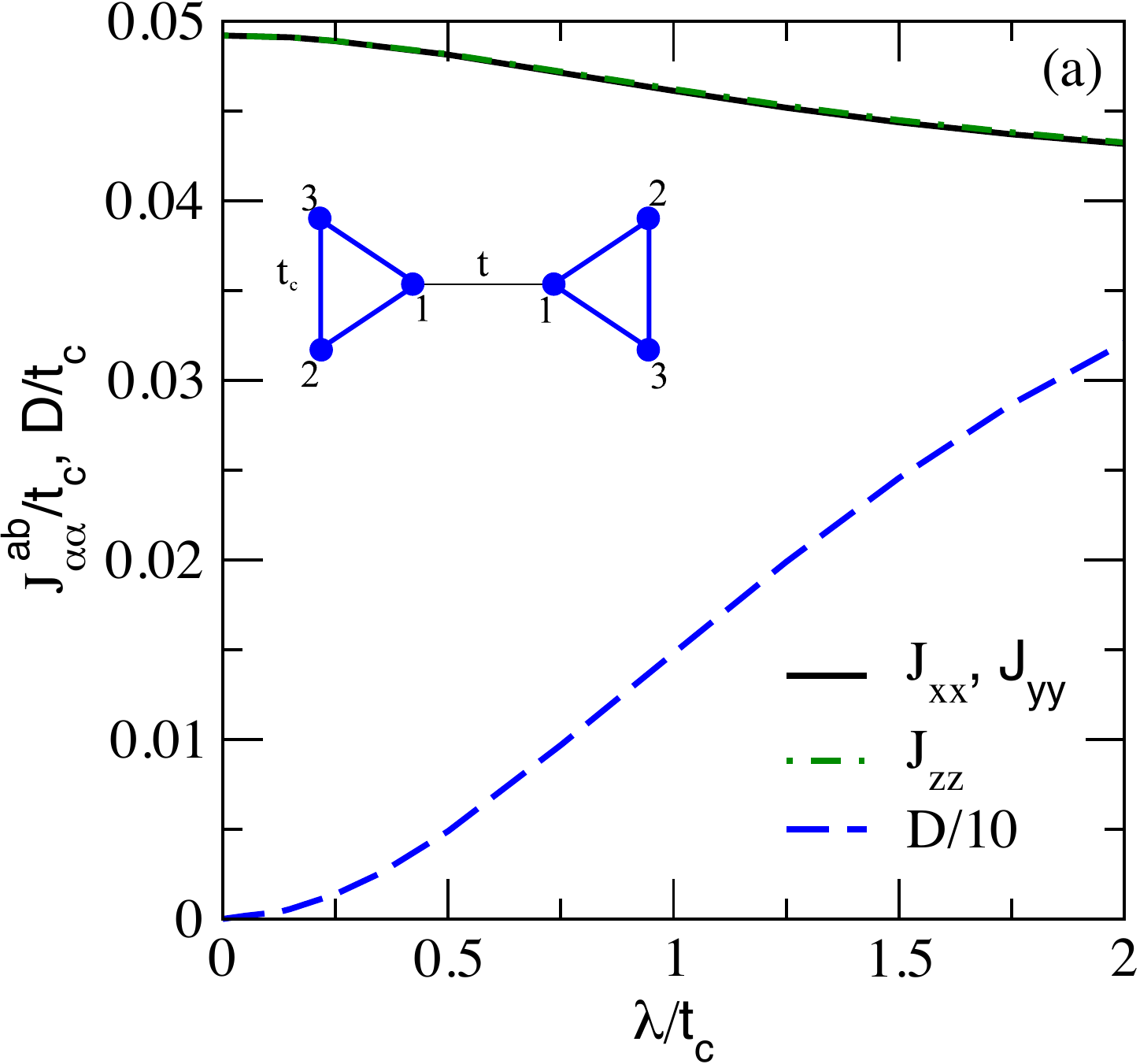} 
   \includegraphics[width=4.25cm, clip=true]{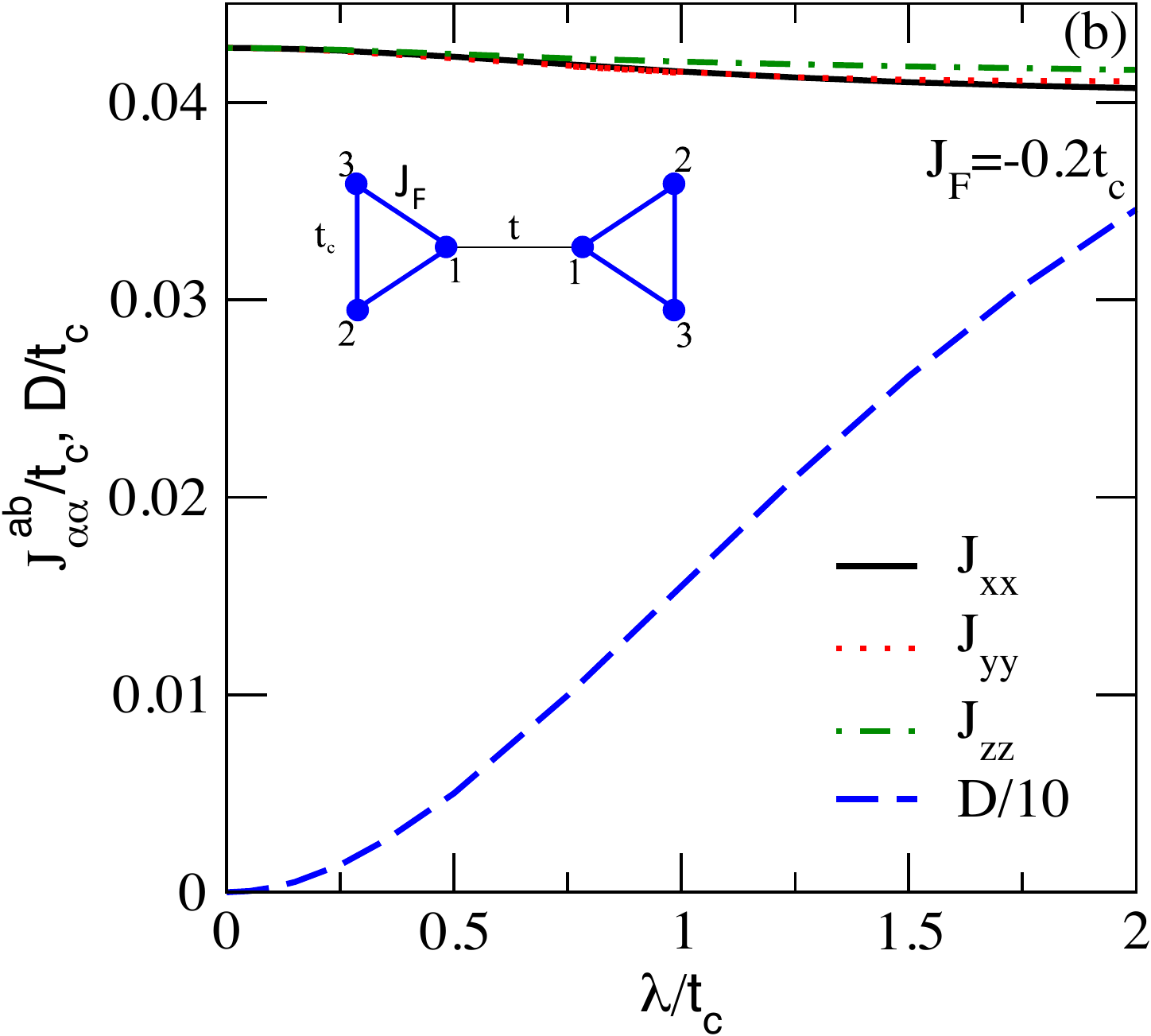} 
   \includegraphics[width=4.35cm, clip=true]{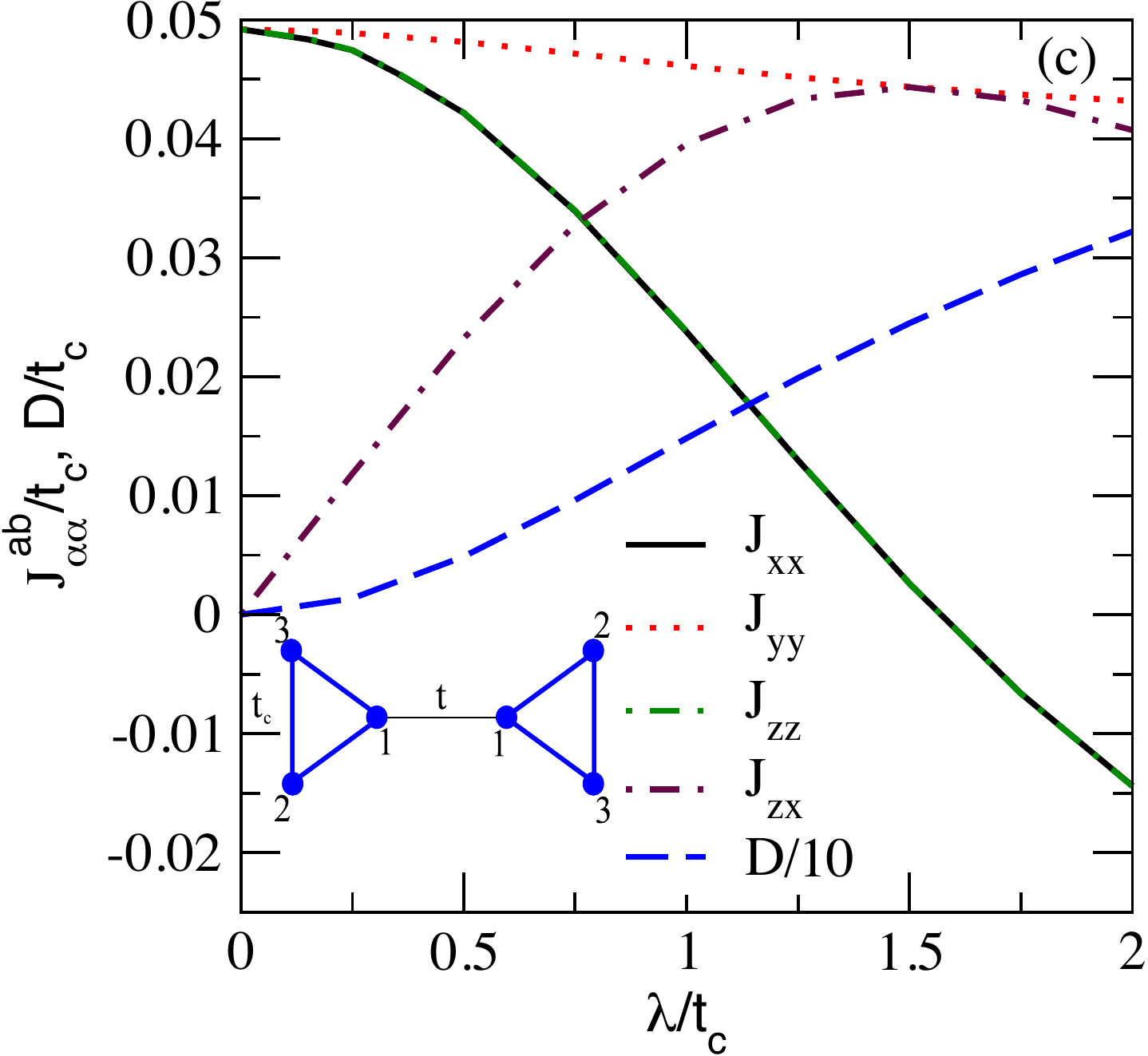}  
    \includegraphics[width=4.2cm, clip=true]{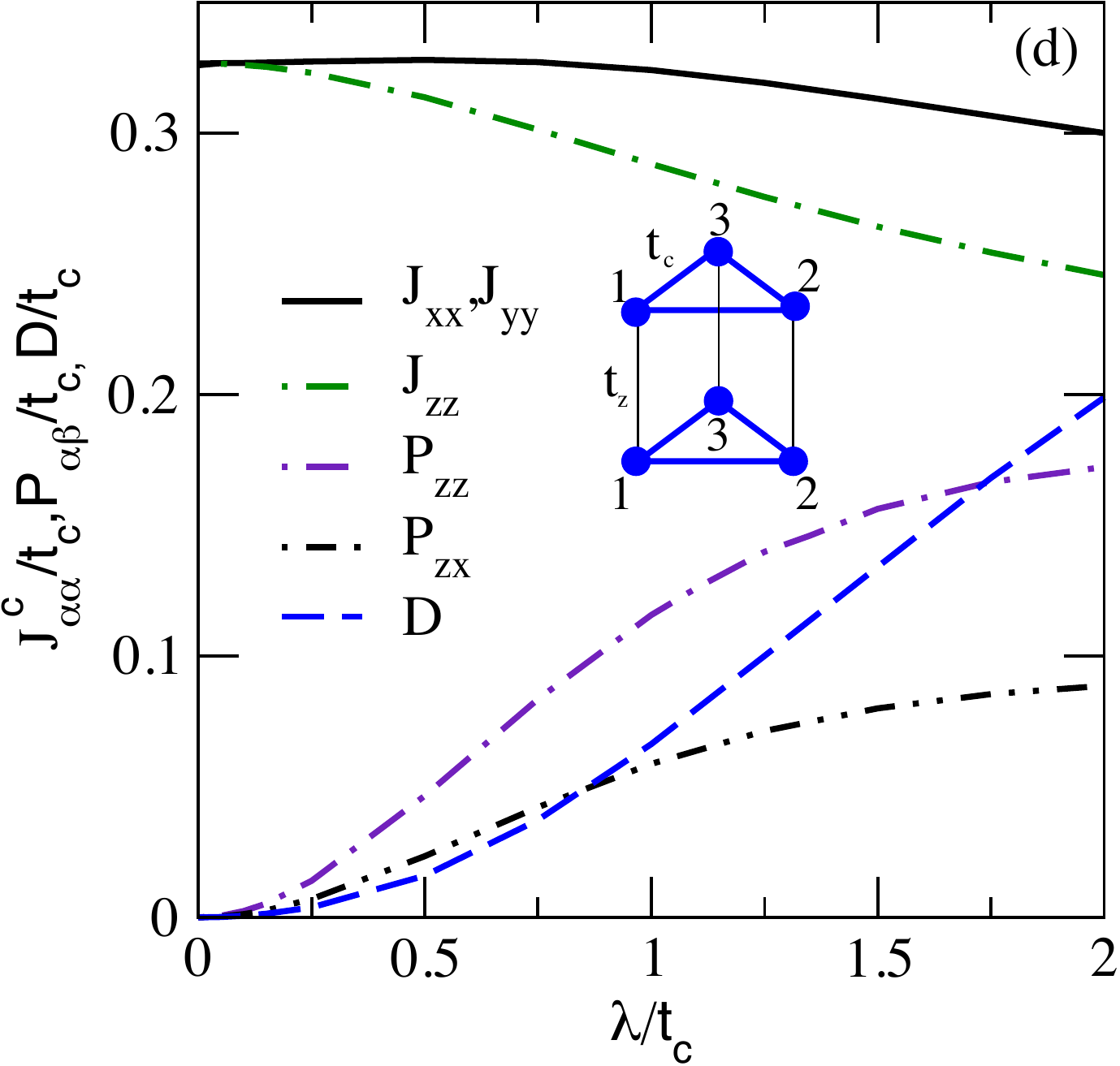} 
   \caption{(Color online) Anisotropic pseudospin exchange coupling, $J_{\alpha\beta}$, and trigonal splitting, $D$, induced by spin-orbit coupling in trinuclear complexes.
We consider the effective interaction between the pseudospins-1 formed in two neighboring
clusters in the $a$-$b$ plane (panels (a)-(c)) and in the $c$-direction of the crystal (panel (d)). In cases (a) and (b) the two molecules 
are related by inversion symmetry whereas in (c) they are related by C$_{2}^{(z)}$ symmetry.  In (d) the two molecules are related
by translational symmetry. We take $U=10t_c$ and the intracluster Heisenberg exchange $J_F=0$ (Hubbard model) except
in case (b) for which, $J_F=-0.2 t_c$. Anisotropic pseudospin exchange interactions
arise when inversion symmetry is broken [cases (c) and (d)] and in the presence of inversion symmetry when $J_F\ne0$ [case (b)].  
%The trigonal splitting $D$ induced by $\lambda$ can tune the system into a $D$-phase for $D \gtrsim J$ 
%occurring at intermediate $\lambda=(0.5-0.75)t_c$ in (a)-(c).  
}
 \label{fig:fig3}
\end{figure}

  The single electron terms are more conveniently expressed in terms of `Bloch' operators on the trimer, $b^\dagger_{mk\sigma}=\sum_j c^\dagger_{mj\sigma}e^{i\phi  kj}/\sqrt{3}$, where  $\phi=2\pi/3$ and $k=0,\pm1$ is the eigenvalue of $L^z_m$   \cite{Amie,powell2015,powell2015SR}. Whence,
 \begin{eqnarray}
 H_{0} &=&-2 t_c\sum_{k\sigma} \cos(k\phi) b^\dagger_{mk\sigma}b_{mk\sigma}
  \nonumber \\
H_{SO} &=& {\lambda_{xy} \over \sqrt{2}} (b^\dagger_{m0\downarrow} b_{m-1\uparrow} -b^\dagger_{m1\downarrow} b_{m0\uparrow} 
  -b^\dagger_{m0\uparrow} b_{m1\downarrow}  
  \nonumber \\
 &+&b^\dagger_{m-1\uparrow} b_{m0\downarrow} )
+{\lambda_z \over 2}  (b^\dagger_{m1\uparrow} b_{m1\uparrow} -b^\dagger_{m1\downarrow} b_{m1\downarrow}
\nonumber\\
 &-&b^\dagger_{m-1\uparrow} b_{m-1\uparrow} +b^\dagger_{m-1\downarrow} b_{m-1\downarrow}). 
\nonumber\\
 \end{eqnarray}
Note that %because of Bloch's   theorem all $k'=k\pm 3n$ with integer $n$  describe equivalent angular momenta. Furthermore 
the trigonal symmetry of the complex implies that the SOC in the  plane of the molecule, $\lambda_{xy}$, need not equal that in the $z$-direction \cite{Amie}. Nevertheless, in some of the numerical work below it is convenient to set  $\lambda=\lambda_{xy}=\lambda_z$ to reduce the parameter space. 
 %({\it i. e.} $k'=k\pm3n$ are equivalent to $k=0,\pm1$). 
% This model differs from models used in the context of transition metal-ionsÉ (Georges review).
% \begin{eqnarray}
%&&H_{U-J_c}={1\over 3}\sum_k (U -2 J_c) n_{k\uparrow}n_{k\downarrow}
%\nonumber \\
%&+&{1 \over 3}\sum_{k,k',k\ne k'} (U -J_c \cos((k-k') \phi)-J_c) n_{k\uparrow}n_{k'\downarrow}
%\nonumber \\
%&+&{1 \over 3 } \sum_{k,k',q \ne 0} (U-J_c \cos((k'-k-q)\phi)-J_c \cos(q\phi) ) c^\dagger_{k\uparrow} c^\dagger_{k'\downarrow} c_{k'-q\downarrow} c_{k+q\uparrow}
%\nonumber \\
%&+& {1 \over 6 } \sum_{k,k',k \ne k'} (J_c \cos((k'-k)\phi)) (n_{k\uparrow}+n_{k\downarrow})
%\nonumber  \\
%&+&{J_c\over 6} \sum_k (n_{k\uparrow} + n_{k\downarrow}).
%\end{eqnarray}
% To keep the discussion as simple as possible we consider a spherically symmetric SOC: $\lambda=\lambda_{xy}=\lambda_z$ and 
%
Spin-orbit coupling can only mix configurations with the 
 same $z$-component of the total orbital momentum of the cluster:  $j$, which  is  conserved (modulo 3) for 
 any values of $\lambda_{xy},\lambda_z$. 
 With no SOC and $t_c>0$ the low lying states are the triplet states   \cite{merino2006}. 
%  $|j=0\rangle={1 \over \sqrt{2}} ( b^\dagger_{m0\uparrow}b^\dagger_{m-1\uparrow}b^\dagger_{m0\downarrow}b^\dagger_{m1\downarrow} 
% - b^\dagger_{m0\uparrow}b^\dagger_{m1\uparrow}b^\dagger_{m0\downarrow}b^\dagger_{m1\downarrow} ) | 0 \rangle $, $|j=1\rangle =b^\dagger_{m0\uparrow}b^\dagger_{m-1\uparrow}b^\dagger_{m0\downarrow}b^\dagger_{m-1\downarrow} |0\rangle$ and $|j=-1\rangle=b^\dagger_{m0\uparrow}b^\dagger_{m1\uparrow}b^\dagger_{m0\downarrow}b^\dagger_{m1\downarrow} |0\rangle$. The SOC couples the $|j=0\rangle$ state above with two $j=0$ configurations:
%$b^\dagger_{m0\uparrow}b^\dagger_{m-1\uparrow}b^\dagger_{m1\uparrow}b^\dagger_{m-1\downarrow} |0\rangle$ and
%$b^\dagger_{m1\uparrow}b^\dagger_{m0\downarrow}b^\dagger_{m-1\downarrow}b^\dagger_{m1\downarrow} |0\rangle$,   
%while it couples the $|j=+1\rangle$ state to only one $j=+1$ configuration, 
%$b^\dagger_{m0\uparrow}b^\dagger_{m-1\uparrow}b^\dagger_{m0\downarrow}b^\dagger_{m-1\downarrow} |0\rangle$,
%and similarly for the $|j=-1\rangle$ state.  Hence, with increasing 
In Fig. \ref{fig:fig2} we show how the cluster spectrum for $\lambda\ne0$, consists on a $|j=0\rangle$ ground state 
and doubly degenerate, $|\pm j\rangle$ states, expected from the combination of time reversal and trigonal ($C_3$) symmetries. At large $U$ the three lowest states 
behave as a localized pseudospin-1. In terms of the pseudospins the effective low energy Hamiltonian for a single complex is thus
$H_{c}^\text{eff}= D ({{\mathcal{ S}}}_m^z)^2$;  $D$ is the trigonal splitting which increases with $\lambda$, cf. Figs. \ref{fig:fig2}, \ref{fig:fig3}. 
%Such a Hamiltonian is also valid for $J_F\neq 0$ as shown in Fig. \ref{fig:fig1}.
%with $\Delta(\lambda)= 0.483 t_c$ for $\lambda=5$ (does $\Delta(\lambda)\rightarrow 0.5$ as $\lambda \rightarrow \infty$??) obtained from Fig. \ref{fig:fig1}. 
%although $J_F$ modifies amplitudes of the wavefunctions as compared to the $J_F=0$.
%the Hubbard term, $H_U$, in orbital space contains the same Coulomb interaction, $U/3$, between electrons in any of the orbitals whereas in
%the Hubbard-Heisenberg model electrons in different orbitals have different Coulomb interaction. 
%(details of the form of the hamiltonians in supplementary?). 
%favoring anisotropic couplings between the pseudospins for the in-plane cluster arrangement in
%contrast to the  pure Hubbard model ($J_F=0$).
 
{\em Exchange interaction between neighboring trinuclear complexes.} We now consider the interaction between two neighboring 
trimers arranged as in the Mo$_3$S$_7$(dmit)$_3$ crystal.  %The Mo$_3$S$_7$(dmit)$_3$ molecules are modeled through Hubbard-Heisenberg trimers containing four electrons as above.  
We analyze the two nearest-neighbor arrangements realized in the crystal. (A) Nearest neighbors  in the $a$-$b$ plane (Fig. \ref{fig:fig1}(a)): a single hopping amplitude, $t$, connects the two molecules: $H_{kin}=-t\sum_{\sigma} \left(c^\dagger_{l1\sigma}c_{m1\sigma}+c^\dagger_{m1\sigma}c_{l1\sigma} \right) $ [cf. inset Fig. \ref{fig:fig3}(a)]. 
(B) Nearest neighbors in the $c$-direction (Fig. \ref{fig:fig1}(b)) there are three hoppings, $t_z$, connecting equivalent orbitals 
 $i$ of the two molecules: $H_{kin}=-t_z\sum_{i\sigma} \left( c^\dagger_{li\sigma}c_{mi\sigma}+c^\dagger_{mi\sigma}c_{li\sigma} \right)$ [cf. inset Fig. \ref{fig:fig3}(d)].  
%We show in Fig. 3 the lowest lying nine 
%eigenstates obtained exactly for the two trimers coupled in the two different ways for different parameters and/or symmetries.  
%These nine eigenstates encode the effective interaction between the two pseudospins arising in the isolated molecules.  
In \Mo first principles  calculations find that the intramolecular hopping is nearly isotropic: $t_z/t=0.87$.

We have calculated the effective exchange coupling between two neighboring clusters via two independent methods: (1) analytically via a canonical transformation to first order in $1/U$ and second order in $H_{SO}$ and $H_{kin}$; and (2) numerically with $H_c$ treated exactly and straightforward second order perturbation theory in $H_{kin}$:
$H^{(2)}_\text{eff}=\sum_{| m^0 \rangle} { H_{kin} |m^0\rangle \langle m^0 | H_{kin} \over 2 E_0(4) - \langle m_0|H_0+H_U+H_{SOC} |m_0\rangle}$,
where $E_0(4)$ is the ground state energy of an isolated cluster with $N=4$ electrons. Note that we  neglect the trigonal splitting ($D$) in the denominator of the numerical perturbation theory. However, comparison with the canonical transformation and the eigenvalues found from exact diagonalisation shows that this approximation is extremely accurate. 
%and does not lead to any qualitative effects, in particular it does not change the structure of the degeneracies discussed below. 
%Virtual excitations induced by hopping between the two clusters involve $N=3$ and $N=5$ electron cluster states. 
%(full expression of $H_{eff}$ in the supplementary section?).

%\\
(A) \emph{Nearest neighbors in the $a$-$b$ plane}. Two neighboring molecules, $l$ and $m$, related by inversion symmetry satisfy: $\lambda_{m,xy}=\lambda_{l,xy}$ and $\lambda_{m,z}=\lambda_{l,z}$. The canonical transformation yields an effective pseudospin model consisting of onsite and nearest neighbour terms \cite{supinfo}:
\begin{eqnarray}
H^\text{eff}_{ab}&=& 
\sum_{\langle lm\rangle\alpha\beta}J^{ab}_{\alpha\beta} {\mathcal S}^\alpha_l {\mathcal S}^\alpha_m 
+ \sum_l \Big\{ D({\mathcal S}^z_l)^2 
\nonumber \\ && 
+ \big[ K_{\pm\pm}  {\mathcal S}_l^+{\mathcal S}_l^+
+   \eta K_{z\pm}{\mathcal S}_l^z {\mathcal S}_l^x 
 \label{eq:istr} 
 +H.c. \big] \Big\}
,
\end{eqnarray}
with $\eta=1$, $J_{\alpha\beta}=J_{\beta\alpha}$, $\alpha,\beta=x,y,z$ and angled brackets imply that the sum is over nearest neighbors only avoiding double counting. The matrix elements calculated from the numerical perturbation theory are in excellent agreement with this model, even for  $\lambda>t_c$. This is not entirely surprising because $\lambda/U$ remains small in this limit. The small residual 
 terms \cite{footnote1} have the symmetry one would expect on extrapolating Eq. (\ref{eq:istr}) to higher orders in $H_{SO}$.

In the pure Hubbard model ($J_F=0$) we find numerically that there are no additional terms at higher order in $1/U$ or $\lambda$, furthermore  
the exchange coupling tensor is diagonal, $J^{ab}_{\alpha\beta}=J^{ab}_{\alpha\alpha}\delta_{\alpha\beta}$, and $K_{z\pm}=K_{\pm\pm}=0$.
% The model consists of two contributions: an on cluster trigonal splitting described by $D$ 
%and the diagonal exchange coupling tensor, $J^{ab}_{\alpha\beta}=J^{ab}_{\alpha\alpha}\delta_{\alpha\beta}$,  
%which describes the exchange coupling between two neighbor pseudospins in the $a$-$b$ plane. 
We plot the dependence of $D$ and $J^{ab}_{\alpha\alpha}$ on $\lambda$ for fixed $U=10t_c$ in Fig. \ref{fig:fig3}(a). 
The exchange couplings only display a very weak anisotropy, $J^{ab}_{xx}=J^{ab}_{yy}\approx J^{ab}_{zz}$,
indicating that the coupling can be described through a standard isotropic Heisenberg model in the presence
of a trigonal splitting induced by SOC. Note  the trigonal splitting, $D$, increases with increasing $\lambda$, and
around $\lambda \approx 0.5t_c$, it becomes comparable to the exchange: $D \approx J^{ab}_{\alpha\alpha}$. 
At large values of  $D\gtrsim J^{ab}_{\alpha\alpha}$, a large-$D$ phase simply given by the tensor product of 
$j=0$ states located on each cluster  is expected.  For instance, in the one-dimensional $S=1$ chain 
a quantum phase transition from the Haldane phase to the large-$D$ phase occurs 
\cite{oitmaa2009,normand2011,langari2013,tzeng2008} for $D/J\sim 0.96-0.971$.

Interestingly, for $J_F \neq 0$  all three diagonal exchange couplings 
of the tensor become different:  $J^{ab}_{zz} > J^{ab}_{yy} > J^{ab}_{xx}$ (Fig. \ref{fig:fig3}(b)).
%(* give some explanation about this behavior based on the form of the cluster hamiltonian *)
Hence, for $J_F\neq0 $ magnetic anisotropies between the pseudospins arise in the
presence of inversion symmetry. The two molecule problem has $C_{2h}$ symmetry once SOC is included. Since all irreducible representations  
of $C_{2h}$ are one-dimensional we should then expect anisotropic couplings between the complexes that lift 
completely the energy level degeneracies present in the isolated trimers (Fig. \ref{fig:fig2}). 
However, these level degeneracies persist in the spectrum of a pair of molecules 
for $J_F=0$ since the in-plane exchanges ($J^{ab}_{xx}=J^{ab}_{yy}$) are equal  in this case.
%, indicating a hidden symmetry in the problem when $J_F=0$.  
Thus,  lowering the symmetry of the Coulomb interaction (\emph{e.g.} $J_F\ne0$) can significantly increase the exchange anisotropy, as in models of transition metal oxides \cite{aharony,perkins2014a}.

Finally we mention that the trigonal splitting, $D$, is only weakly affected by $J_F$ (compare Figs. \ref{fig:fig3}(a) and (b)).
The off-diagonal exchange, $J_{\alpha\beta}$, and non-pseudospin-conserving $K_{\alpha\beta}$ terms are non-zero, but remain small ($\sim10^{-4} t_c$), for the parameters explored in Fig. \ref{fig:fig3}(b).

It is important to understand what happens in the absence of an inversion center relating neighboring complexes. A particularly interesting case is if the two molecules are related by a rotation of $\pi$ about a z-axis bisecting the two molecules (C$_{2}^{(z)}$ symmetric). In the absence of SOC the C$_{2}^{(z)}$ symmetric and inversion symmetric models are identical. But the pseudovectorial nature of angular momenta implies that for a pair of molecules, $l$ and $m$, related by  C$_{2}^{(z)}$ symmetry $\lambda_{m,xy}=-\lambda_{l,xy}$ and $\lambda_{m,z}=\lambda_{l,z}$.
With these relations the canonical transformation again yields an effective pseudospin Hamiltonian  given by Eq. (\ref{eq:istr}) \cite{supinfo}; but in contrast to the inversion symmetric case we now have $\eta=(-1)^{l+1}$ and 
%\begin{eqnarray}
%H^{eff}_{ab}&=&\sum_l \Big\{ D ({\mathcal S}^z_l)^2+[{1 \over 2} \{(-1)^{l+1}K_{z\pm}({\mathcal S}_l^z{\mathcal S}_l^++{\mathcal S}_l^z {\mathcal S}^-_l)
%\nonumber \\
%&+& K_{\pm\pm}{\mathcal S}^+_l {\mathcal S}^+_l+H.c.\}] \Big\}
%+\sum_{\langle lm\rangle\alpha\beta}J^{ab}_{\alpha\beta} {\mathcal S}_{l\alpha} {\mathcal S}_{m\beta}.
%\label{eq:c2z}
%\end{eqnarray}
%In this case: 
$J^{ab}_{\beta\alpha}=-J^{ab}_{\alpha\beta}$ for $\alpha\ne\beta$. Furthermore $J^{ab}_{xx}=J^{ab}_{zz} \neq J^{ab}_{yy}$ and $J^{ab}_{xy}=J^{ab}_{yz}=0$.  The latter equality 
is straightforward since the hamiltonian is real when written in the orbital basis.
Because the off-diagonal exchange is antisymmetric these terms are equivalent to a  Dzyaloshinskii-Moriya interaction, ${\bf D}_{lm}\cdot\bm{{\mathcal S}}_l\times\bm{{\mathcal S}}_m$,  \cite{DM1,DM2} with $D_{lm}^y=J^{ab}_{zx}/\sqrt{2}$ and $D_{lm}^x=D_{lm}^z=0$.

  Anisotropies are large in  the $C_2^{(z)}$ model  [Fig. \ref{fig:fig3}(c)],  with $J^{ab}_{zz}$ even becoming ferromagnetic for sufficiently large SOC: $\lambda > 1.7t_c$. %Due to breaking inversion symmetry there is an off-diagonal exchange coupling, $J^{ab}_{zx} \ne 0$, which leads to a symmetric interaction between the $x,z$ components of the two pseudospins.  This
$J^{ab}_{zx}$ can become as
large as the diagonal $J^{ab}_{\alpha\alpha}$ exchange couplings.  The non-pseudospin-conserving terms induced by SOC at O($\lambda^2$) are found to be small
compared to the other contributions [$K_{\alpha\beta}/t_c\sim10^{-4}-10^{-3}$ for the parameters explored in Fig. \ref{fig:fig3}(c)].  The trigonal splitting is again large, with $D\approx J^{ab}_{yy}>J^{ab}_{\alpha\alpha}$,  (with $\alpha=x,z$) at moderate values of $\lambda\approx 0.45t_c$.

%\\
(B) {\it Nearest neighbors in the $c$-direction}  display a tube arrangement [Figs. \ref{fig:fig1} and \ref{fig:fig3}(d) inset] with the two clusters are related by 
translational symmetry (no inversion symmetry center). The canonical transformation yields the effective pseudospin model:
\begin{eqnarray}
H^\text{eff}_{c}&=&D\sum_l({\mathcal{S}}_{l}^z)^2
+\sum_{\langle lm\rangle\alpha\beta}J^c_{\alpha\beta} \mathcal{S}_{l}^{\alpha} \mathcal{S}_{m}^{\beta} 
\nonumber \\&& 
+ \sum_{\langle lm\rangle\alpha\beta} P_{\alpha\beta} \mathcal{S}_{l}^{\alpha} \mathcal{S}_{l}^{\beta} \mathcal{S}_{m}^{\alpha} \mathcal{S}_{m}^{\beta},
\label{eq:AKLT}
\end{eqnarray}
%where apart from the trigonal splitting, $D$, and the anisotropic quadratic spin couplings, $J_{\alpha\beta}$,  
where anisotropic biquadratic couplings, $P_{\alpha\beta}=P_{\beta\alpha}$, obey $P_{zz}=2P_{zx}=2P_{zy}$ and $P_{xx}=P_{yy}=P_{xy}=0$, these relations are also confirmed numerically.  The dependence of 
the non-zero couplings on $\lambda$ is shown in Fig. \ref{fig:fig3}(d). For $J_F=0$ the off-diagonal couplings $J^c_{\alpha\beta}= 0$ for $\alpha \neq \beta$,
whereas the diagonal terms behave as $J^c_{xx} =J^c_{yy} > J^c_{zz}$.  %The only biquadratic terms which are non-zero in the presence of SOC are the diagonal $P_{zz}$ and  the off-diagonal $P_{zx}$ contributions which are not independent: $P_{zz}=2P_{zx}$.  
Note that the isotropic 
version of  model (\ref{eq:AKLT}),  \emph{i.e.}, $J^c_{\alpha\beta}=J^c\delta_{\alpha\beta}, P_{\alpha\beta}=P\delta_{\alpha\beta}$ 
and $D=0$ is just the bilinear-biquadratic model:  $H=J^c{\bf S}_l \cdot {\bf S}_m +P({\bf S}_l\cdot{\bf S}_m)^2$, which 
becomes the Affleck-Kennedy-Lieb-Tasaki (AKLT) model for $P/J^c=1/3$, which has the valence bond solid ground state
which is in the Haldane phase \cite{AKLT}.  
%Introducing these states in Eq. (\ref{eq:hameff}) we find (supplementary?):
%\begin{eqnarray}
%H_{eff}^{(2)}=t^2 \sum_{\gamma_l,\gamma_m} \sum_{\sigma,\sigma'} \sum_{\mu_l,\nu_m,\mu'_l,\nu'_m} A_{\gamma_l}(3,\mu_l) A_{\gamma_m}(5,\nu_m) A^*_{\gamma_l}(3,\mu'_l) A^*_{\gamma_m}(5,\nu'_m) 
%\nonumber \\
%{c^\dagger_{l1\sigma}c_{m1\sigma}|3, \mu_l \rangle  |5,\nu_m \rangle \langle 5,\nu'_m | \langle 3,\mu'_l|
%c^\dagger_{m1\sigma'}c_{l1\sigma'} \over \Delta \epsilon(3,\gamma_l;5, \gamma_m) }
%\nonumber \\
%+t^2 \sum_{\gamma_l,\gamma_m} \sum_{\sigma,\sigma'} \sum_{\mu_m,\nu_l,\mu'_m,\nu'_l} A_{\gamma_l}(5,\nu_l) A_{\gamma_m}(3,\mu_m) A^*_{\gamma_m}(3,\mu'_m) A^*_{\gamma_l}(5,\nu'_l) 
%\nonumber \\
%{c^\dagger_{m1\sigma}c_{l1\sigma}|3, \mu_m \rangle  |5,\nu_l \rangle \langle 5,\nu'_l | \langle 3,\mu'_m|
%c^\dagger_{l1\sigma'}c_{m1\sigma'} \over \Delta \epsilon(3,\gamma_m;5, \gamma_l) }
%\label{eq:2nd}
%\end{eqnarray}
%where $\Delta \epsilon(3,\gamma_l;5,\gamma_m)=2 E_0(4)-(E_{\gamma_l}(3)+E_{\gamma_m}(5))$ denote the excitation energy in going 
%from the ground state of the  two isolated clusters to having 3 electrons in the $l$ cluster and 5 electrons in the $m$ cluster. 
\begin{figure}%[h]
  \includegraphics[width=5cm]{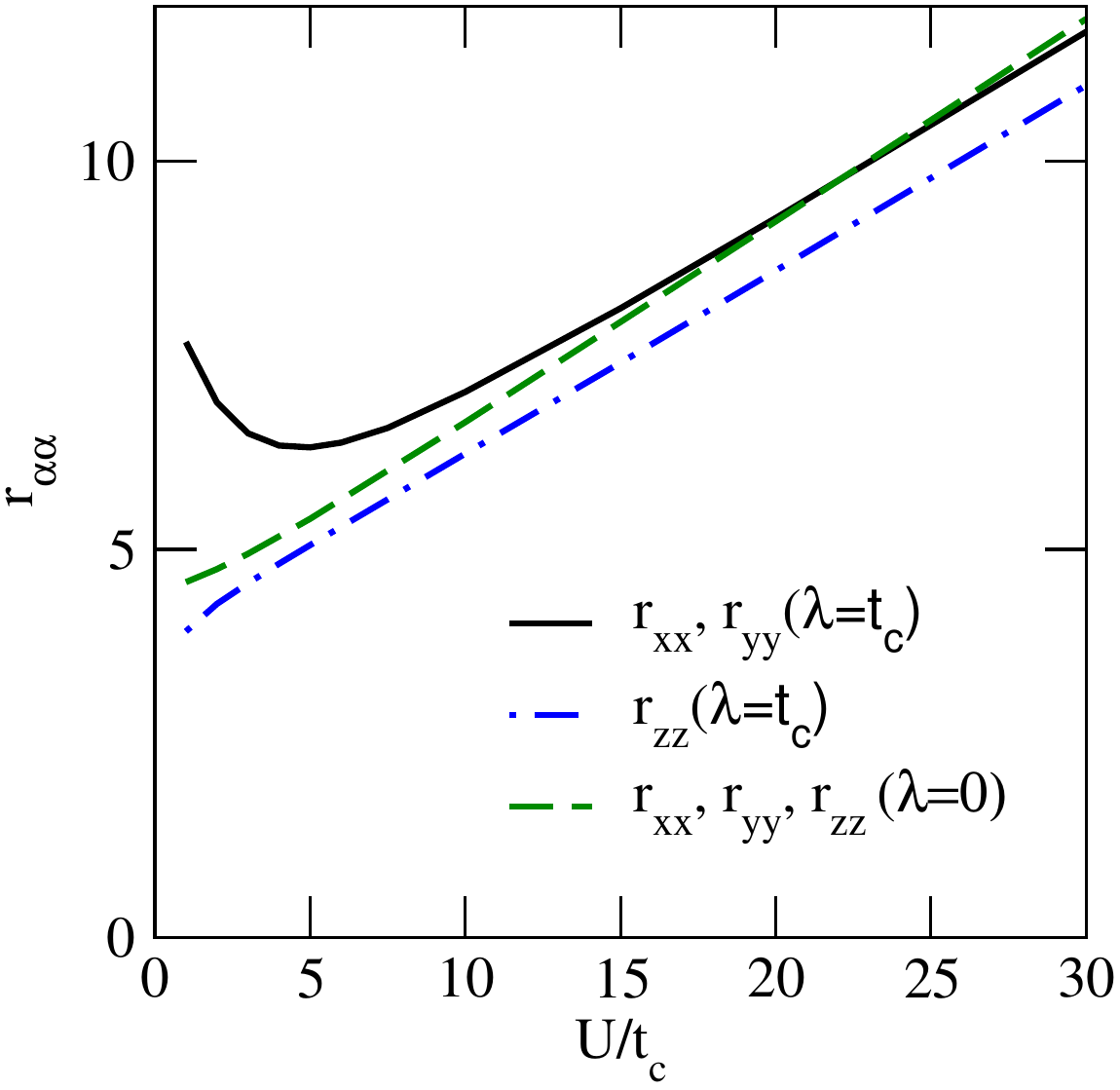}
   \caption{(Color online) Effective one-dimensionality arising from electron correlations in trinuclear complexes. 
   %The $U$-dependence of the ratios of the nearest neighbor spin exchange couplings 
   %in the $c$-direction to the nearest-neighbor exchange couplings in the $a$-$b$ plane, 
   %$r_{\alpha\alpha}=J^c_{\alpha\alpha}/J^{ab}_{\alpha\alpha}$, with $\alpha=x,y,z$ is shown.
  As the on-site Coulomb repulsion ($U$) increases spatial anisotropy of the exchange coupling, $r_{\alpha\alpha}=J^c_{\alpha\alpha}/J^{ab}_{\alpha\alpha}$, is enhanced driving the crystal to a quasi-one-dimensional system.}
 \label{fig:fig4}
\end{figure}

The in-plane  isotropy of the interlayer exchange ($J_{xx}=J_{yy}$, \emph{etc}.) arises from  the trigonal symmetry of the molecular packing (tube). In contrast to the symmetry of the in-plane packing (dumbbell) motif which allows  $J_{xx}\ne J_{yy}$. 

{\em Quasi-one-dimensional pseudospin-1 model for trinuclear complexes.}
Comparing the magnitude of the  diagonal exchange couplings in the $a$-$b$ plane, $J^{ab}_{\alpha\alpha}$
[Fig. \ref{fig:fig3}(a)-(c)], with the pseudospin exchange couplings in the $c$-direction, $J^{c}_{\alpha\beta}$  
[Fig. \ref{fig:fig3}(d)],  one  finds significant spatial anisotropy:  $r_{\alpha\alpha}=J^c_{\alpha\alpha}/J^{ab}_{\alpha\alpha}\gg1$.  Furthermore, $r_{\alpha\alpha}$ increases rapidly with $U$, see Fig. \ref{fig:fig4}.
Indeed $J^{ab}_{\alpha\alpha}\rightarrow0$ for all $\alpha$ as $U\rightarrow\infty$ independent of the other parameters in the Hamiltonian  \cite{supinfo}. In contrast all $J^{c}_{\alpha\alpha}$ remain finite as $U\rightarrow\infty$ in the absence of fine tuning of  $\lambda_{xy}$, $\lambda_z$ and $t_c$ \cite{supinfo}.
%This behavior indicates that electronic correlations {\it enhance} the effective couplings along the $c$-direction
%making the crystal effectively behave as a one-dimensional system is spite of the fact that the crystal is nearly 
%isotropic \cite{jacko2015}. 
Thus even though the underlying electronic structure is highly isotropic ($t_z/t=0.87$ \cite{jacko2015}) strong electronic correlations result in a quasi-one-dimensional spin model.
%The large enhancement of the ratios occur regardless of the strength of the SOC. 
This emergent quasi-one-dimensionality is a consequence of the fact that nearest neighboring complexes are connected by three hopping integrals to the molecule along the $c$-direction of the crystal but only one hopping integral connects neighbor molecules in the $ab$-planes. 
Interference effects \cite{footnote2}  among the possible O($t^2$) paths contributing to the exchange couplings make the ratio, $r_{\alpha\alpha}$,
a non-universal number that increases with $U$ \cite{supinfo}.  Hence, in the strong coupling limit, the 
effective model for trinuclear complexes consists on weakly coupled antiferromagnetic 
pseudospin-1  chains described by model (\ref{eq:AKLT}).

{\em Conclusions and outlook.} Our work shows how antiferromagnetic pseudospin-1 chains arise along the $c$-direction in  crystals of trinuclear complexes such as Mo$_3$S$_7$(dmit)$_3$ and its selenated analogs.
With no SOC, these chains can be modeled through an isotropic $S=1$ Heisenberg model
whose ground state is in the Haldane phase  \cite{janani2014a,nourse}. An important question is whether the Haldane phase is stable 
to the weak exchange coupling between neighbor chains arranged in the hexagonal geometry.  Previous numerical work \cite{sengupta2014}
has shown how the Haldane phase is unstable to the interchain exchange coupling when $r=J^c/J^{ab}>3$ in 
the hexagonal geometry of Mo$_3$S$_7$(dmit)$_3$ crystals. Since, $r_{\alpha\alpha}\gtrsim5$ for all $\alpha$, for even moderate $U$, we expect the Haldane phase to be the ground 
state of the crystal for weak SOC. 

Turning on SOC introduces a trigonal splitting, $D$, as well as anisotropic spin exchange 
couplings.  Thus, increasing SOC can drive the crystal from a topological Haldane phase to a trivial `$D$-phase'  consisting of 
the tensor product of  $j=0$ states on each molecule. Hence, a quantum phase
transition \cite{weisel2014} may be induced 
in the family of materials based on Mo$_3$S$_7$(dmit)$_3$ crystals by substituting, say, Mo$\rightarrow$W \cite{Amie} or S$\rightarrow$Se \cite{winter2012}, which effectively increases the SOC. Intriguingly, although several selenated analogs of \Mo have been synthesized \cite{guschin2013},  little is known about their magnetic properties.

\emph{Acknowledgements.}
J.M.  acknowledges financial support from: (MAT2012-37263-C02-01, MAT2015-66128-R) MINECO/FEDER, UE.
Work at the University of Queensland was supported by the Australian Research Council (FT130100161, DP130100757  and DP160100060).


\begin{thebibliography}{99}
\bibitem{supinfo} Analytic expressions for the parameters in Eqs. (\ref{eq:istr})-(\ref{eq:AKLT}) to first order in $1/U$ and second order in $H_{SO}$ and $H_{kin}$ are given in the Sup. Info.
\bibitem{footnote1} From our calculations we estimate that residual terms are at most of ${\cal O}(10^{-4}t_c)$.
\end{thebibliography}
\end{document}

\title{Supplementary information for ``Topological quantum phase transition driven by anisotropic spin-orbit coupling in trinuclear organometallic coordination crystals''}
\author{J. Merino}
\affiliation{Departamento de F\'isica Te\'orica de la Materia Condensada, Condensed Matter Physics Center (IFIMAC) and
	Instituto Nicol\'as Cabrera, Universidad Aut\'onoma de Madrid, Madrid 28049, Spain}
\author{A. C. Jacko}
\affiliation{School of Mathematics and Physics, The University of Queensland,
	Brisbane, Queensland 4072, Australia}
\author{A. L. Khosla}
\affiliation{School of Mathematics and Physics, The University of Queensland,
	Brisbane, Queensland 4072, Australia}
\author{B. J. Powell}
\affiliation{School of Mathematics and Physics, The University of Queensland,
	Brisbane, Queensland 4072, Australia}

\begin{abstract}
 
\end{abstract}

\maketitle

\section{Complexes related by inversion symmetry}\label{sect:inv}

For a nearest neighbor complexes in the $a$-$b$ plane related by inversion symmetry the effective Hamiltonian to leading order in $1/U$ and  second order in $H_{SO}$ and $H_{kin}$ is given by Eq. (2) of the main text. To leading order in $J_c=J_F+4t_c^2/U$ and  second order in $H_{SO}$ and $H_{kin}$ the parameters in Hamiltonian (2) for this situation are:
\begin{subequations}
	\begin{eqnarray}
	J_{\alpha\beta}&=&J_{\beta\alpha},\\
	J_{xx}&=&
		\frac{4t_g^2}{9U}  \left[1
		-\frac{1}{48} \left(\frac{\lambda_z}{t_c}\right)^2
		\right]
		+\frac{2t_g^2}{81}\left(\frac{4}{U}+\frac{J_F}{ t_c^2}\right)
		\left[ 1  
		+ \frac{5 }{144}\left(\frac{\lambda_z}{t_c}\right)^2 
		-\frac{1}2\left(\frac{\lambda_{xy}}{t_c^2}\right)^2 
		\right],
	\\
	J_{yy}&=& 
		\frac{4t_g^2}{9U}  \left[1
		-\frac{1}{48} \left(\frac{\lambda_z}{t_c}\right)^2
		\right]
		+\frac{2t_g^2}{81}\left(\frac{4}{U}+\frac{J_F}{ t_c^2}\right)\left[ 1  
		+ \frac{5 }{144}\left(\frac{\lambda_z}{t_c}\right)^2 
		-\frac{1}{3}\left(\frac{\lambda_{xy}}{t_c^2}\right)^2
		\right],	
		\label{Jyy_g_pp}
	\\
	J_{zz}&=&   
	 \frac{4t_g^2}{9U}
	+\frac{2t_g^2}{81}\left(\frac{4}{U}+\frac{J_F}{ t_c^2}\right)
		\left[1
		+\frac{  1 }{12 } \left(\frac{\lambda_z}{t_c}\right)^2
		-\frac{7 }{12 } \left(\frac{\lambda_{xy}}{t_c^2}\right)^2
	\right],
	\\
	J_{xz}&=&
		\frac{1}{\sqrt{2}}
		\frac{t_g^2}{486}\left(\frac{1}{U}+\frac{J_F}{4  t_c^2}\right)\frac{\lambda_{xy}\lambda_z}{t_c^2},
	\\
	J_{yz}&=&0,\\
	J_{xy}&=&0,\\
	K_{z\pm}
	&=&\sqrt{2}\left[
	\frac{t_g^2}{81}
	\left(\frac{1}{6t_c^3}+\frac{11}{6t_c^2U}+\frac{11J_F}{24t_c^4}\right)\right]\lambda_{xy}\lambda_z,
	\label{Kzp_g_pp}
\\
	K_{\pm\pm}&=& 
	\frac{t_g^2}{81}\left(\frac{2}{3U}+\frac{ J_F  }{6t_c^2}\right)\left(\frac{\lambda_{xy}}{t_c^2}\right)^2,
	\label{Kpp_g_pp}
\\
D
&=&
 \left[\frac{1}{6 t_c} +\frac{1}{6U}+\frac{J_F}{24t_c^2}\right]\lambda_{xy}^2 
-\left[\frac{1}{12 t_c} -\frac{1}{6U} -\frac{J_F}{24t_c^2}\right]\lambda_z^2
\notag\\&&
-\frac{t_g^2}{81}\left[\frac{5}{24t_c^3}\lambda_z^2 +  
\left(\frac{43}{72t_c^2U} + \frac{43J_c}{288t_c^4} \right)\lambda_z^2 -
\left(\frac{1}{3 t_c^2U} + \frac{J_F}{12 t_c^4}\right) \lambda_{xy}^2\right] 
\label{Delta_g_pp}.
	\end{eqnarray}
\end{subequations}

\section{Complexes related by $C_2^{(z)}$}

For complexes related by a rotation by $\pi$ about the $z$-axis then $\lambda_{xy}$ must have equal magnitudes but opposite signs on the two complexes. For a nearest neighbor complexes in the $a$-$b$ plane the effective Hamiltonian to leading order in $1/U$ and  second order in $H_{SO}$ and $H_{kin}$ is again given by Eq. (2) of the main text. To leading order in  $J_c=J_F+4t_c^2/U$ and  second order in $H_{SO}$ and $H_{kin}$ one finds that $J_{yy}$, $K_{z\pm}$, $K_{\pm\pm}$, and $D$ are given by Eqs. (\ref{Jyy_g_pp}),  (\ref{Kzp_g_pp}),  (\ref{Kpp_g_pp}),  and (\ref{Delta_g_pp}) respectively, but
\begin{subequations}
	\begin{eqnarray}
	J_{xx}&=&\frac{4t_g^2}{9U}  \left[1-\frac{1}{48} \left(\frac{\lambda_z}{t_c}\right)^2
	-\frac{4}{9} \left(\frac{\lambda_{xy}}{t_c}\right)^2
	\right]
	+\frac{2t_g^2}{81}\left(\frac{4}{U}+\frac{J_F}{ t_c^2}\right)\left[ 1  
	+ \frac{5 }{144}\left(\frac{\lambda_z}{t_c}\right)^2 
	-\frac{5}{18}\left(\frac{\lambda_{xy}}{t_c^2}\right)^2 
	\right],
\\
	J_{zz}&=&   \frac{4t_g^2}{9U}\left[1-\frac49\left(\frac{\lambda_{xy}}{t_c}\right)^2\right]
	+\frac{2t_g^2}{81}\left(\frac{4}{U}+\frac{J_F}{t_c^2}\right)\left[1
	+\frac{  1 }{12 } \left(\frac{\lambda_z}{t_c}\right)^2
	-\frac{13 }{36 } \left(\frac{\lambda_{xy}}{t_c^2}\right)^2
	\right],
\\
	J_{xz}=-J_{zx}&=&
	\frac{1}{\sqrt{2}} 
	\frac{2t_g^2}{81}\left\{\left(\frac{76}{3U}+\frac{J_F}{3t_c^2}\right)
		\frac{\lambda_{xy}}{t_c}
			+\left(\frac{155}{36U}+\frac{11J_F}{144t_c^2}\right)
	\frac{\lambda_{xy}\lambda_z}{t_c^2}
	\right\}.
	\end{eqnarray}
\end{subequations}

\section{Effective Hamiltonian along the chains}
For complexes stacked along the $c$-axis, related by translational symmetry, the low-energy effective Hamiltonian is given by Eq. (4) of the main text to leading order in $1/U$ and  second order in $H_{SO}$ and $H_{kin}$. To leading order in  $J_c=J_F+4t_c^2/U$ and  second order in $H_{SO}$ and $H_{kin}$
\begin{subequations}
	\begin{eqnarray}
	J_{xx}=J_{yy}
	&=&\frac{4t_z^2}{3U}  
	+\frac{t_z^2}{9}\left\{\frac{2}{ t_c}\left[ 1  
	+ \frac{1}{18}\left(\frac{\lambda_z}{t_c}\right)^2
	- \frac{5}{36}\left(\frac{\lambda_{xy}}{t_c}\right)^2 \right]
	+\left(\frac{4}{U}+\frac{J_F}{t_c^2}\right)\left[ 1  
	+ \frac16\left(\frac{\lambda_z}{t_c}\right)^2 
	-\frac{29}{72}\left(\frac{\lambda_{xy}}{t_c}\right)^2
	\right]
	\right\},
	\\
	J_{zz}&=&   
	\frac{4t_z^2}{3U}   
	+\frac{t_z^2}{9}\left\{\frac{2}{ t_c}\left[ 1  
	+ \frac{1}{36}\left(\frac{\lambda_z}{t_c}\right)^2
	- \frac{5}{36}\left(\frac{\lambda_{xy}}{t_c}\right)^2 \right]
	+\left(\frac{4}{U}+\frac{J_F}{t_c^2}\right)\left[ 1  
	+ \frac1{12}\left(\frac{\lambda_z}{t_c}\right)^2 
	-\frac{4}{9}\left(\frac{\lambda_{xy}}{t_c}\right)^2
	\right]
	\right\},
	\\	
	P_{zz}&=&   
	\frac{t_z^2}{9}\left[\frac{1}{2t_c} + \frac{3}{U} + \frac{3J_F}{4t_c^2} \right]\left(\frac{\lambda_z}{t_c}\right)^2,
	\\
	D
	&=&
	-\frac{t_z^2}{81}\left\{\left[\frac{7}{2t_c}+\frac{43}{2U}
	\right]
	\left(\frac{\lambda_z}{t_c}\right)^2
	- \left(\frac{3}{2U}+\frac{3J_F}{8t_c^2}\right)\left(\frac{\lambda_{xy}}{t_c}\right)^2
	\right\}.
	\end{eqnarray}
\end{subequations}